\documentclass[preprint2]{emulateapj}

\shorttitle{Vela Shrapnel B}
\shortauthors{Yamaguchi and Katsuda}

\begin{document}

\title{Suzaku Spectroscopy of Vela Shrapnel B
}

\author{H. Yamaguchi\altaffilmark{1}}
\email{hiroya@crab.riken.jp}


\author{S. Katsuda\altaffilmark{2,3}}

\altaffiltext{1}{RIKEN (The Institute of Physical and Chemical Research), 
  2-1 Hirosawa, Wako, Saitama 351-0198, Japan}
\altaffiltext{2}{Department of Earth and Space Science, Osaka University, 
  1-1 Machikaneyama, Toyonaka, Osaka 560-0043, Japan}
\altaffiltext{3}{NASA Goddard Space Flight Center, 
  Greenbelt, MD 20771, USA}

\begin{abstract}

We present the X-ray observation of Vela shrapnel~B with the XIS on 
board the {\it Suzaku} satellite.  The shrapnel is one of several ejecta 
fragment-like features protruding beyond the primary blast wave shock 
front of the Vela supernova remnant.  The spectrum of shrapnel~B is
well-represented by a single-temperature thin-thermal plasma in a
non-equilibrium ionization state.  The elemental abundances of O, Ne,
and Mg are found to be significantly higher than the solar values,
supporting that shrapnel~B originates from supernova ejecta.  
The abundances of O, Ne, and Mg relative to Fe are enhanced above their
solar values, while that of Si relative to Fe are at their
solar values.  This abundance pattern is similar to that in
shrapnel~D, except that the enhancements of the lighter elements are
less prominent, suggesting more extensive mixing with the
interstellar medium (ISM) in shrapnel~B.  The contribution of the ISM
is considered to be larger at the trailing region, because the
absolute abundances of some elements there are depleted relative to
those at the shrapnel's head.

\end{abstract}


\section{Introduction}
\label{sec:introduction}

Supernova (SN) explosions are prominent phenomena through which 
heavy elements are released into interstellar space. 
The ejected materials often constitute fragments, or ejecta knots, 
although their creation mechanism has not been well-understood. 
In fact, such fragments of ejecta have been discovered from several 
supernova remnants (SNRs): e.g., Tycho (Decourchelle et al.\ 2001), 
Cas~A (Laming \& Hwang 2003; Fesen et al.\ 2006), 
Pup~A (Winkler \& Kirshner 1985; Katsuda et al.\ 2008), 
and G292.0+1.8 (Park et al.\ 2007).  As mentioned below, \object{the
  Vela SNR} is also one of the most interesting SNRs from which some
candidates of ejecta fragments can be observed.

The Vela SNR is considered to be the remnant of a Type II-P SN explosion 
of the progenitor star with the mass of $<$~25$M_{\odot}$ (Gvaramadze 1999). 
The age of the SNR has been estimated to be $\sim 1.1\times 10^4$~yr from 
the spin down rate of PSR B0833--45 (the Vela pulsar: Taylor et al.\ 1993), 
which is associated with the SNR (e.g., Weiler \& Sramek 1988). 
Since this SNR has the largest angular size ($\sim 8^{\circ}$: 
Aschenbach et al.\ 1995; Lu \& Aschenbach 2000) among observable SNRs, 
it is a very ideal target for studying the detailed structure of an
old SNR.

Aschenbach et al.\ (1995) observed the entire Vela SNR in the {\it ROSAT} 
all-sky survey, and discovered six isolated fragment-like features, named 
``shrapnels'' A to F, which have overrun the primary blast wave shock 
boundary. The opening angles of the shrapnels' bow-shock features suggest 
them to be supersonically moving in tenuous ambient matter. 
The symmetry axis of each shrapnel traces back to the center of 
the SNR, which is very close to the Vela pulsar. Such a configuration 
suggests that the shrapnels are associated with the fossil material of 
the SN explosion; they had probably traveled behind the blast wave for 
a long time and passed its front at the late stage when the blast wave 
had decelerated. If they are really the clumps of the SN ejecta, it is 
expected that the chemical compositions of the shrapnels are abundant in 
heavy elements. Therefore, some of the shrapnels have been studied to 
reveal the metal abundances using several recent X-ray missions.

Shrapnel~A is the most well-investigated one. Tsunemi et al.\ (1999a) 
observed the shrapnel with {\it ASCA}, and found a strong Si K-shell 
emission line with relatively weak emission from other elements. Then, 
the {\it Chandra} and {\it XMM-Newton} observations confirmed that the 
shrapnel is significantly over-abundant in Si, while the other elements, 
i.e., O, Ne, Mg, and Fe, have solar or sub-solar abundances 
(Miyata et al.\ 2001; Katsuda \& Tsunemi 2006: hereafter KT06). 
They thus concluded that shrapnel~A originated from a deep (Si-rich) layer 
of the progenitor star. The relative abundances among the heavy elements 
were found to be almost uniform in the entire shrapnel.

On the other hand, the {\it ASCA} spectra of shrapnel~D, which is the
brightest and largest among the six shrapnels, indicated 
over-abundances of O, Ne, and Mg (Tsunemi et al.\ 1999b). Using {\it
  XMM-Newton}, Katsuda \& Tsunemi (2005; hereafter KT05) confirmed
that the abundance ratios of O-Ne-Mg to Fe are significantly larger
than solar, which is consistent with the view that this shrapnel is a
fragment of the ejecta from a shallow layer of the progenitor.

Interestingly, several bright ejecta knots have been discovered in the 
northern part of the Vela SNR by the recent {\it XMM-Newton} observations 
(Miceli et al.\ 2008). 
They suggested that these knots would be ``new'' shrapnels hidden inside 
the Vela main shell by the projection effect. The relative abundances of 
these knots were found to be quite similar to those observed in shrapnel~D; 
O, Ne, and Mg were much more abundant than those of the other elements.

Shrapnel~B is another candidate of the ejecta fragment discovered with 
{\it ROSAT} all-sky survey, and the second brightest next to 
shrapnel~D (Aschenbach et al.\ 1995). Although it was observed with 
{\it ASCA}, the elemental abundances could not be determined due to 
the poor photon statistics (Tsunemi et al.\ 1999b). 
In this paper, therefore, we investigate the chemical and physical 
properties of shrapnel~B, by analyzing high quality spectral data
obtained with {\it Suzaku}.  In \S2, we introduce the observation and
data reduction procedure.  The results of our analysis are presented
in \S3.  Finally, we develop the discussion and summarize our findings
in \S4 and \S5, respectively.  We assume the distance to the Vela SNR
to be 250~pc, following Cha et al.\ (1999).  The errors quoted in 
the text and tables are at the 90\% confidence level, and the error
bars in the figures are for 1$\sigma$ confidence, unless otherwise
noted.

\section{Observation and Data Reduction}
\label{sec:observation}

\begin{figure}
\includegraphics[scale=.45]{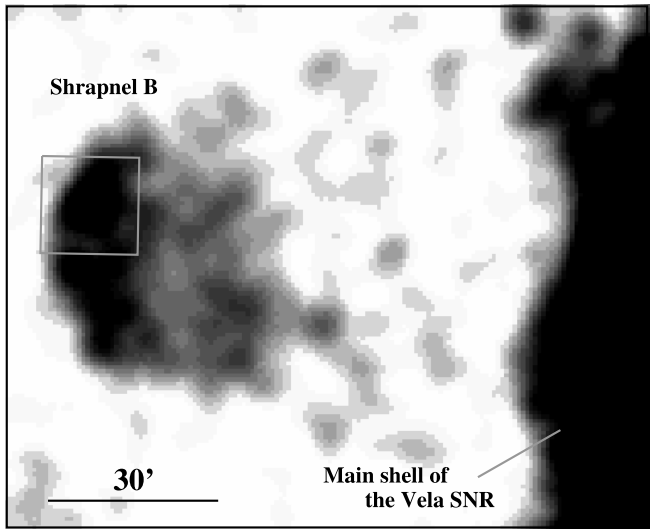}
\caption{Surface brightness map of Vela shrapnel~B in the 0.5--1.3~keV band 
  obtained with {\it ROSAT} PSPC. North is up and east is to the left. 
  The scale is logarithmic. The field of view of the XIS is shown 
  by a gray solid square. 
  \label{fig:rosat}}
\end{figure}

The 0.5--1.3~keV {\it ROSAT} PSPC image of shrapnel~B
(Figure~\ref{fig:rosat}) shows a clear bow shock structure.  A part of
Vela shrapnel~B was observed by {\it Suzaku} on 2006 November  
5--7 (Observation ID = 501085010). The field of view (FOV) of the 
observation is shown in the image of the entire region of 
shrapnel~B. 

In this paper, we concentrate on the data obtained with the X-ray 
Imaging Spectrometer (XIS: Koyama et al.\ 2007) on board {\it Suzaku}. 
The XIS consists of four X-ray charge coupled devices (CCD) at the foci of 
the four X-Ray Telescopes (XRT; Serlemitsos et al.\ 2007). All four XRTs are 
co-aligned to image the same region of the sky. The half-power diameter (HPD) 
of the XRT is $\sim 2'$, independent of the X-ray energy. Three of the CCDs 
(XIS0, 2, and 3) are Front-Illuminated (FI) sensors and the fourth (XIS1) is 
Back-Illuminated (BI).  The latter has superior sensitivity 
in the 0.2--1.0~keV band with a significantly improved energy resolution 
compared to previous X-ray CCD cameras.

The XIS were operated in the normal full-frame clocking mode with 
spaced-row charge injection (SCI) technique (Uchiyama et al.\ 2008). 
For the following data reduction, the HEADAS software of version 6.5 
was used. We employed the revision 2.0 data products, but 
reprocessed the data using the \texttt{xispi} software and 
the \texttt{makepi} files of 20080905. These files include the 
latest calibration results for the charge transfer inefficiency (CTI) 
and gain.  We cleaned the reprocessed data following 
the standard screening
criteria\footnote{http://heasarc.nasa.gov/docs/suzaku/processing/criteria\_xis.html}
and obtained an effective exposure time of $\sim$58~ksec. 
Although all four sensors were active at the time of the observation, 
the XIS2 became inoperable soon after the use of the SCI
technique. The energy scale of the XIS2 for  
the SCI-on mode was therefore not well calibrated (Uchiyama et al.\
2008) and we do not use the data obtained by the XIS2 in the
subsequent analysis.

\section{Analysis and Results}
\label{sec:analysis}

\subsection{Image}
\label{ssec:image}

\begin{figure}
\includegraphics[scale=.7]{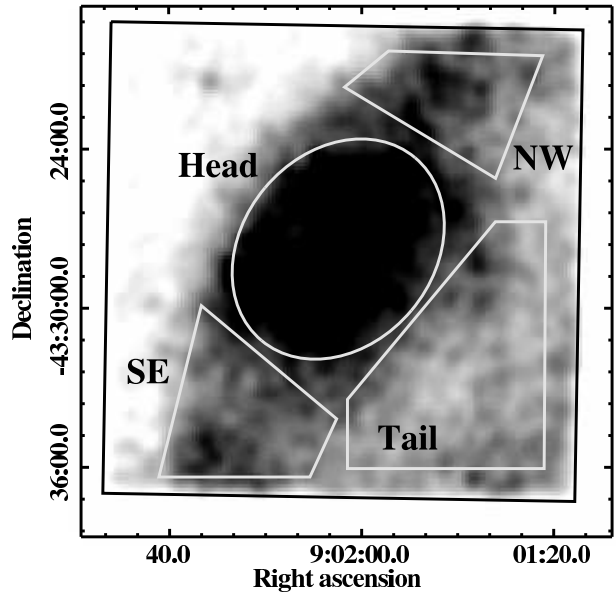}
\caption{{\it Suzaku} XIS image of Vela shrapnel~B in the 0.5--2.0~keV 
  band, with J2000.0 coordinates. The data from the XIS0, 1, and 3 are 
  combined. The vignetting effect is corrected after subtraction of 
  the non X-ray background. The image is smoothed with a Gaussian 
  kernel of $\sigma = 33''$ and shown on a linear scale. 
  The field of view of the XIS is indicated by the black square. 
  The source regions used for the spectral analyses are shown by 
  the solid lines. 
  \label{fig:image}}
\end{figure}

Figure~\ref{fig:image} shows the XIS image in the 0.5--2.0~keV band. 
No significant diffuse emission is found above 2~keV, which suggests 
that the shrapnel is dominated by soft X-rays. 
An image of the non X-ray background (NXB) in the same energy band was 
obtained using the \texttt{xisnxbgen} software and subtracted from the 
raw image. The NXB data were constructed from the night-Earth database 
by sorting on the geomagnetic cut-off rigidity. 
The NXB-subtracted image was then divided by the exposure map constructed 
with the \texttt{xissim} software to correct for vignetting.

In this image, we can see the clear ridge structure of the shrapnel. 
The diffuse emission 
is the brightest at the center of the FOV and extends to the northwest 
and southeast. 
It is also extended to the opposite side of the rim (toward the main 
shell of the Vela SNR), although its surface brightness there is lower
than in the other regions. Similar trailing features are
found in shrapnels~A and D (Miyata et al.\ 2001; KT05; KT06).

\subsection{Spectrum}
\label{ssec:spectrum}

For a quantitative study of the physical properties, we extracted 
the XIS spectra from several regions indicated with solid lines in 
Figure~\ref{fig:image}---hereafter, ``Head'', ``Tail'', ``NW'', and
``SE''.  
We analyzed the spectra using XSPEC version 11.3.2, with 
the response matrix files (RMF) and ancillary response files (ARF) 
made using with the \texttt{xisrmfgen} and \texttt{xissimarfgen} software.

\subsubsection{Background subtraction}
\label{sssec:bgd}

Although there is a source-free region at the east edge of the FOV, it is 
too small to obtain a background spectrum with sufficient photon statistics. 
Therefore, we used ``blank sky'' data to estimate the background emission. 
The most suitable blank sky observation for our data is the offset
observation for  
RX~J0852--4622 (Obs.\ date = 2005 December 23, Obs.\ ID = 500010020),
whose aim point (RA, Dec) = 
(9$^{\rm h}$00$^{\rm m}$31$^{\rm s}$, --43$^{\circ}$28$'$36$''$) is
only $\sim 4.4$ degrees from Vela shrapnel~B.  
We applied the same data-screening criteria as we did for the shrapnel~B 
data, obtaining an exposure time of $\sim 59$~ksec. To minimize the 
uncertainty due to the background subtraction, the background spectra were 
taken from the same detector coordinates as the source regions.

\begin{figure}
\includegraphics[scale=.35]{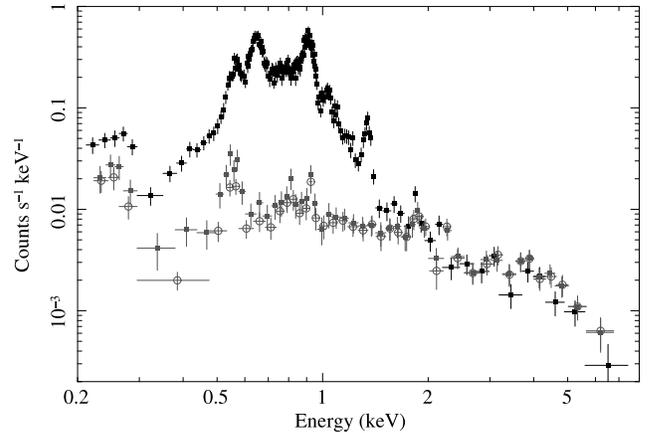}
\caption{Comparison of the source and background spectra. 
  Black with filled squares represents the XIS1 spectrum of Head region
  with the non X-ray background subtracted. 
  The corresponding blank sky spectra before and after the efficiency 
  correction are represented with gray with filled squares and open 
  circles, respectively (see text for details). 
  \label{fig:compare}}
\end{figure}

It is known that the energy resolution of the XIS is gradually 
degrading due to radiation damage while in orbit (Koyama et al.\ 2007). 
Also, the efficiency for the detection of soft X-rays has decreased
because of the build-up of the contaminating material on the optical
blocking filters (OBF) of the XIS due to out-gassing from the
satellite.  Since these effects vary with the time, we examined the
differences in the XIS performance for the shrapnel~B and blank
sky observations.  We found that the difference in energy resolution
is negligible, at $\sim 60$~eV and $\sim 55$~eV (FWHM) at 0.5~keV for  
the shrapnel~B and blank sky data, respectively. On the other hand,
the column density of the accumulated contaminant (at the center of the
CCDs) during each  
observation shows some differences as shown in Table~\ref{tab:contami}. 
For the XIS0 and 1, the extinction column densities at the time 
of the source observation are twice as large as those for the blank sky, 
so there should be taken properly into account.  
We thus applied the following procedure before the background subtraction: 
(1) The NXB data from the same detector regions were subtracted from 
both of the source and blank sky spectra. (2) The NXB-subtracted 
spectra of the blank sky were corrected by multiplying the ratios 
between the efficiencies during the source and blank sky observations. 
For example, we show in Figure~\ref{fig:compare} the result for the 
Head region. The black and gray data points with filled squares are 
the NXB-subtracted XIS1 spectra from the source and blank sky data, 
respectively. After the efficiency correction, the blank sky spectrum 
is given as the gray crosses with open circles. The count rate of the 
soft X-rays (below $\sim 0.6$~keV) was significantly reduced, 
while that of the hard X-rays was not changed much.

\begin{table}
\begin{center}
\caption{Column densities of the contaminant built up on the optical
  blocking filters of the XIS.\tablenotemark{a}
  \label{tab:contami}}
\begin{tabular}{lcccc}
\tableline\tableline
 Observation (Date)  & Contaminant  & XIS0  & XIS1  & XIS3  \\
\tableline
 Vela shrapnel B & C ($10^{18}$~cm$^{-2}$) &  3.19  &  4.14  &  5.75  \\ 
 (2006-11-05)    & O ($10^{17}$~cm$^{-2}$) &  5.32  &  6.89  &  9.59  \\ 
\tableline
 RX J0852--4622 offset & C ($10^{18}$~cm$^{-2}$) &  1.53  &  2.20  &  4.45  \\ 
 (2005-12-23)          & O ($10^{17}$~cm$^{-2}$) &  2.55  &  3.66  &  7.41  \\ 
  \tableline
\end{tabular}
\tablenotetext{a}{Values for the center positions of the CCDs, 
estimated by the \texttt{xiscontamicalc} software.}
\end{center}
\end{table}

\begin{figure*}
\plottwo{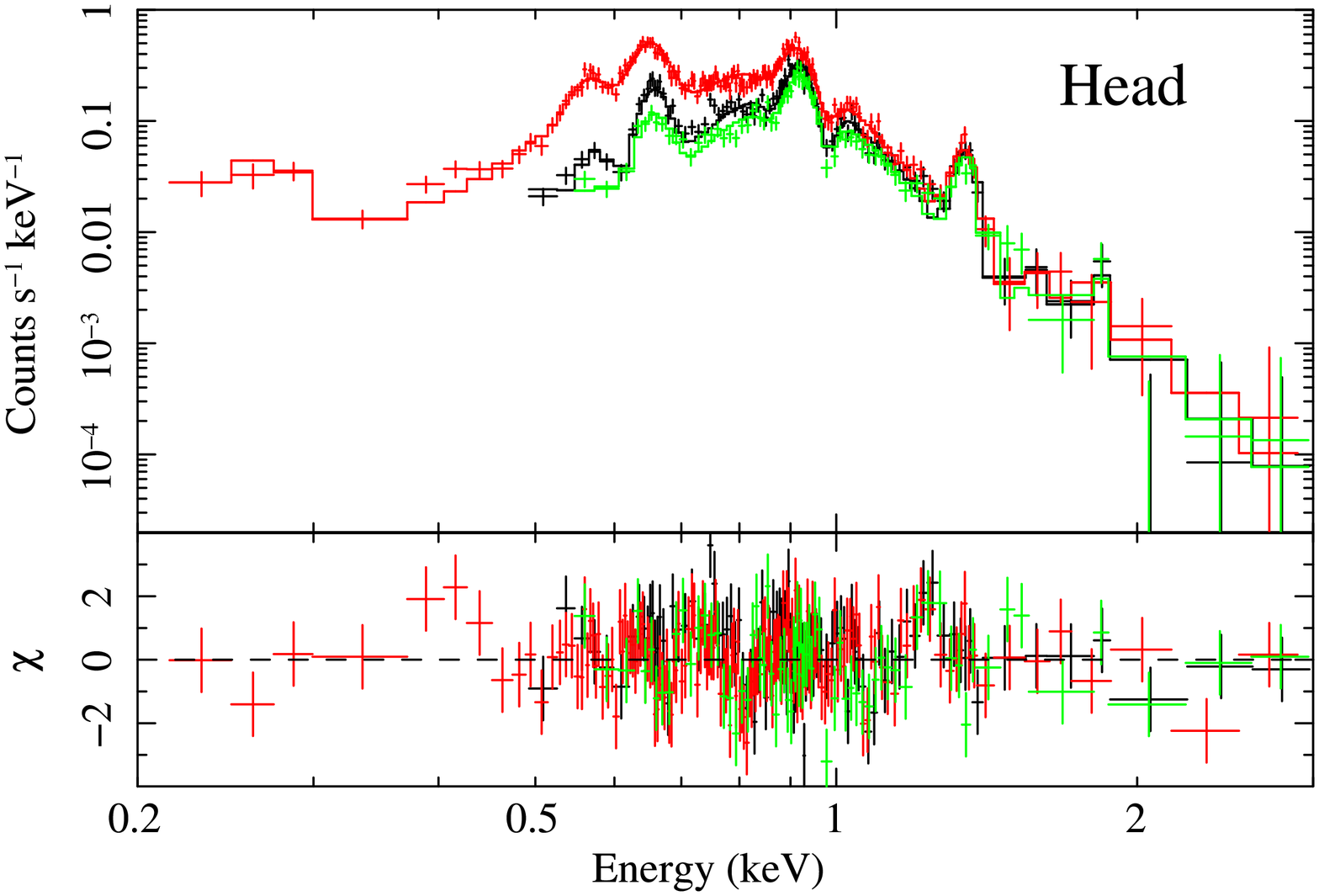}{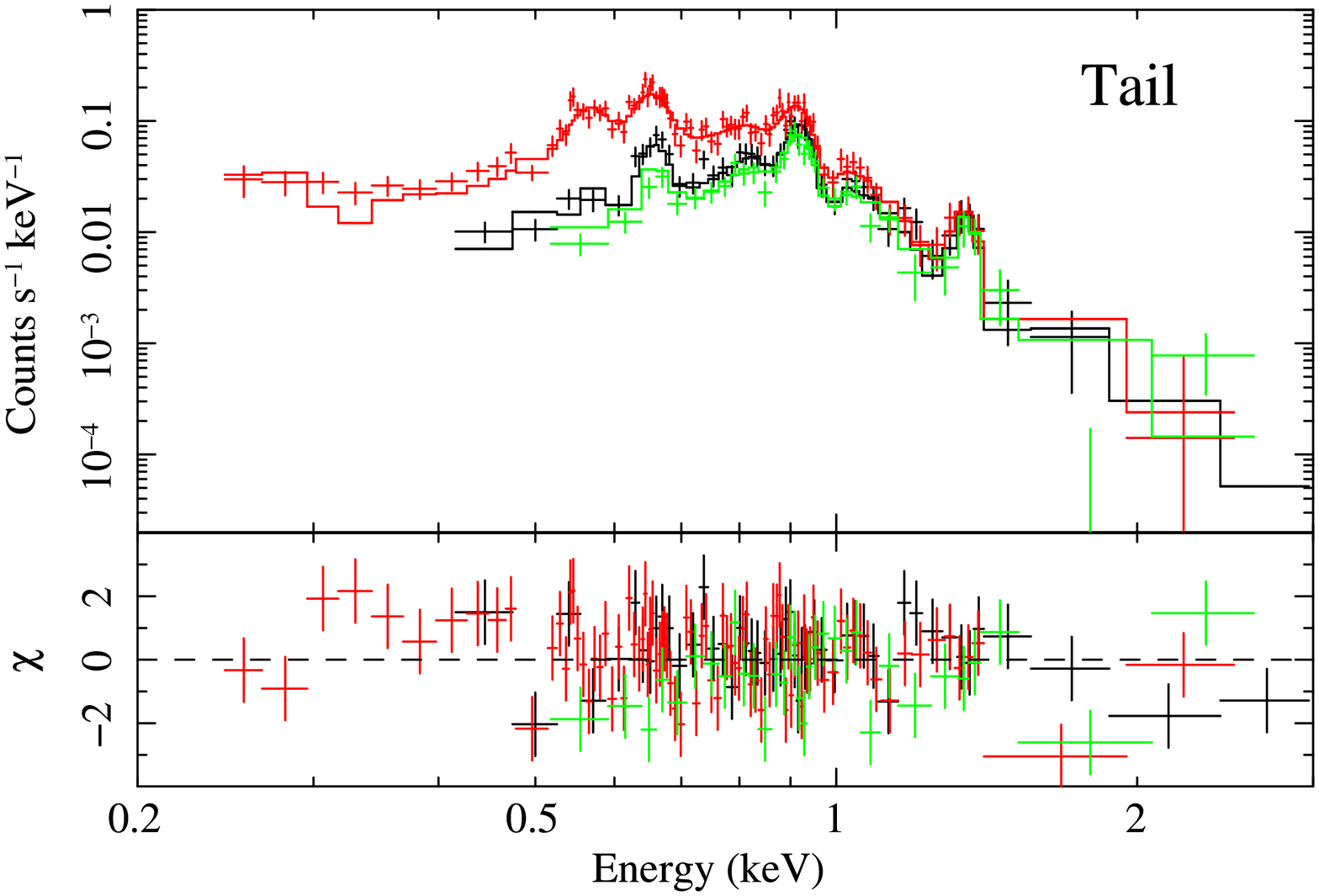}
\plottwo{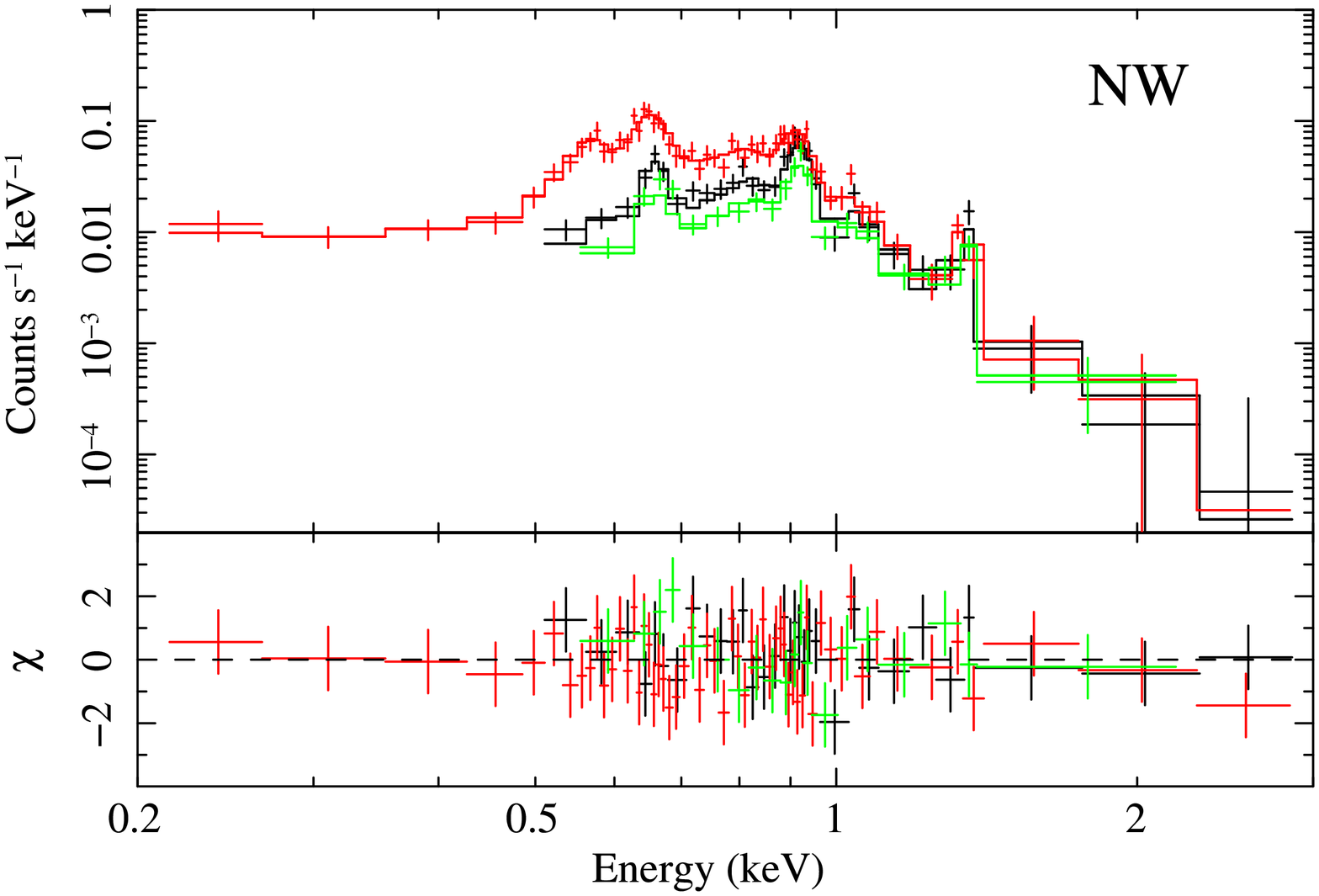}{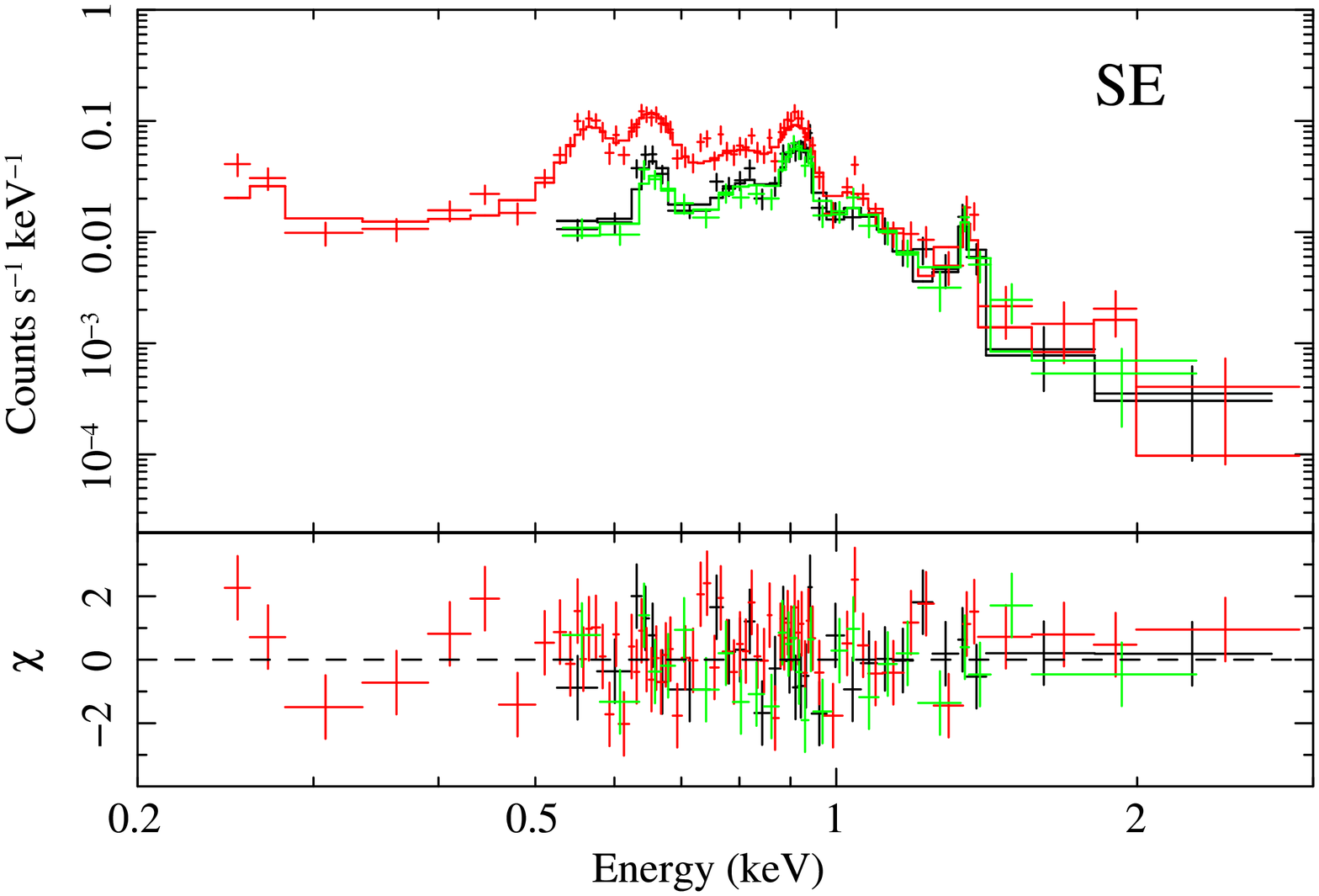}
\caption{Background-subtracted XIS spectrum of each region. 
  The black, red, and green data points represent XIS0, 1, and 3 spectra, 
  respectively. The solid lines show the best-fit NEI models. 
  \label{fig:spectrum}}
\end{figure*}

The background-subtracted spectrum of each region was obtained as shown 
in Figure~\ref{fig:spectrum}.  Several K-shell emission 
line blends, including OVII ($\sim 0.57$~keV), OVIII ($\sim 0.65$~keV), 
NeIX ($\sim 0.91$~keV), NeX ($\sim 1.02$~keV), and MgXI ($\sim 1.34$~keV)
were clearly resolved in each spectrum, clearly indicating origin in
the optically thin-thermal plasma of the shrapnel. 


\subsubsection{Head region}
\label{sssec:head}

\begin{table*}
\begin{center}
\caption{Best-fit spectral parameters of Vela shrapnel B.\label{tab:fit}}
\begin{tabular}{lcccc}
\tableline\tableline
  Parameter & Head  & Tail  & NW  & SE  \\
  \tableline
  $N_{\rm H}$ ($10^{20}$~cm$^{-2}$)  &  1.3 (0.6--2.3)  &  
    1.3 (fixed)  &  1.3 (fixed)  &  1.3 (fixed) \\
  $kT_e$ (keV)  &  0.61 (0.55--0.68)  &  0.69 (0.61--0.81)  &  
    0.53 (0.43--0.66)  &  0.75 (0.58--0.98) \\
  O (solar)  &  1.7 (1.3--2.2)  &  0.77 (0.63--0.94)  &  
    1.4 (0.9--1.9)  &  1.3 (0.9--2.1) \\
  Ne (solar)  &  4.4 (3.5--5.8)  &  1.8 (1.5--2.3)  &  
    3.7 (2.3--6.7)  &  3.2 (2.2--5.2) \\
  Mg (solar)  &  2.9 (2.3--4.0)  &  1.3 (1.0--1.9)  &  
    3.5 (2.0--5.3)  &  2.6 (1.6--4.3) \\
  Si, S (solar)  &  1.5 (0.7--2.5)  &  1.9 (1.2--2.9)  &  
    3.4 (1.2--7.2)  &  4.1 (2.3--5.7) \\
  Fe, Ni (solar)  &  1.4 (1.1--1.8)  &  0.79 (0.63--0.98)  &  
    1.4 (0.8--2.6)  &  1.1 (0.8--1.9) \\
  $n_et$ ($10^{10}$~cm$^{-3}$~s)  &  4.2 (3.5--5.1)  &  
    5.1 (3.6--8.5)  &  3.4 (2.4--5.8)  &  2.5 (1.7--3.6) \\
  EM\tablenotemark{a} ($10^{16}$~cm$^{-5}$) &  3.6 (2.7--4.9)  &
    2.4 (2.0--2.9)  &  2.7 (1.4--5.0)  &  2.0 (1.1--2.9)  \\
  $L_{\rm X}$\tablenotemark{b} ($10^{30}$~ergs~s$^{-1}$) &
    16.2  &  7.9  &  4.0  &  5.4  \\ 
\tableline
  $\chi ^2$/d.o.f.  & 370/318  & 208/168  & 91/102  & 149/120  \\
\tableline
\end{tabular}
\tablenotetext{a}{Emission measure (EM = $\int n_e n_p ~dl$), where $n_e$, 
  $n_p$, and $dl$ are the electron and proton densities and the plasma 
  depth, respectively.}
\tablenotetext{b}{X-ray luminosity in the 0.2--2.0~keV band.}
\end{center}
\end{table*}

We first fitted the spectra of the Head region with a model for an
absorbed optically thin-thermal plasma in collisional ionization
equilibrium (CIE) (VAPEC: Smith et al.\ 2001). The absorption column density 
($N_{\rm H}$), electron temperature ($kT_e$), and emission measure (EM) 
were treated as free parameters. The elemental abundances relative to the 
solar values (Anders \& Grevesse 1989) of O, Ne, Mg, Si, and Fe were also 
allowed to vary freely. The abundances of S and Ni were 
linked to those of Si and Fe respectively, while those of the other
elements were fixed to the solar values. 
Since the absolute gain of the XIS has a small uncertainty of $\pm 5$~eV 
(Koyama et al.\ 2007; Uchiyama et al.\ 2008), we allowed for a small 
offset in the photon-energy to the pulse-height gain relationship. 
With this model, we obtained a best-fit with $kT_e = 0.26 \pm 0.01$~keV 
and $\chi ^2$/d.o.f.\ = 428/319.

Although the spectra were relatively well reproduced by the CIE model, 
we also fitted a non-equilibrium ionization plasma model 
(VNEI with NEIvers 2.0: Borkowski et al.\ 2001).
The ionization timescale ($n_et$), where $n_e$ and $t$ are electron 
density and the time since the plasma was heated, was an additional 
free parameter in these models.  Given that the value of $\chi
^2$/d.o.f.\ was significantly reduced to 370/318 (F-test probability
was less than $10^{-10}$), we considered that the NEI
model better represents the Head spectra. 
The best-fit parameters and model are given in Table~\ref{tab:fit}, 
and Figure~\ref{fig:spectrum} ({\it top left}), respectively. 
A relatively large disagreement between the data and model was found 
around 1.2~keV, which can be attributed to Fe-L emission lines missing in 
the plasma code (e.g., Brickhouse et al.\ 2000). 
The above fitting process required a gain offset of 4.4~eV only for 
the XIS1, which is within the allowable range. 
Although we also tried to fit the spectra by fixing the abundances of 
C and N to that of O, the result did not change significantly. 
The Ne and Mg abundances are enhanced by a factor of a few above the 
solar values. Although it is less prominent, the O is also over-abundant 
than the solar at a 3$\sigma$ confidence level. 

\begin{figure*}
\plottwo{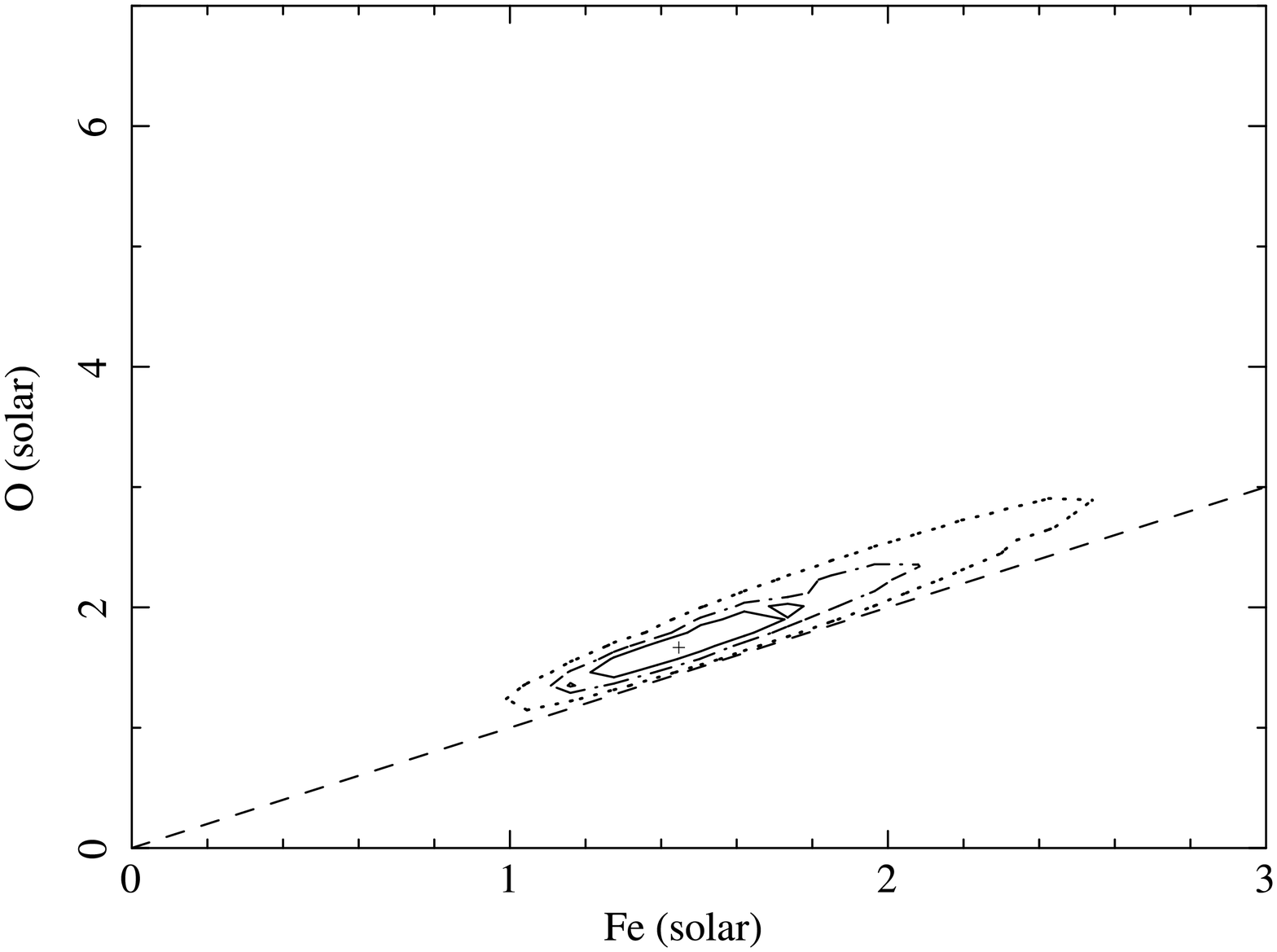}{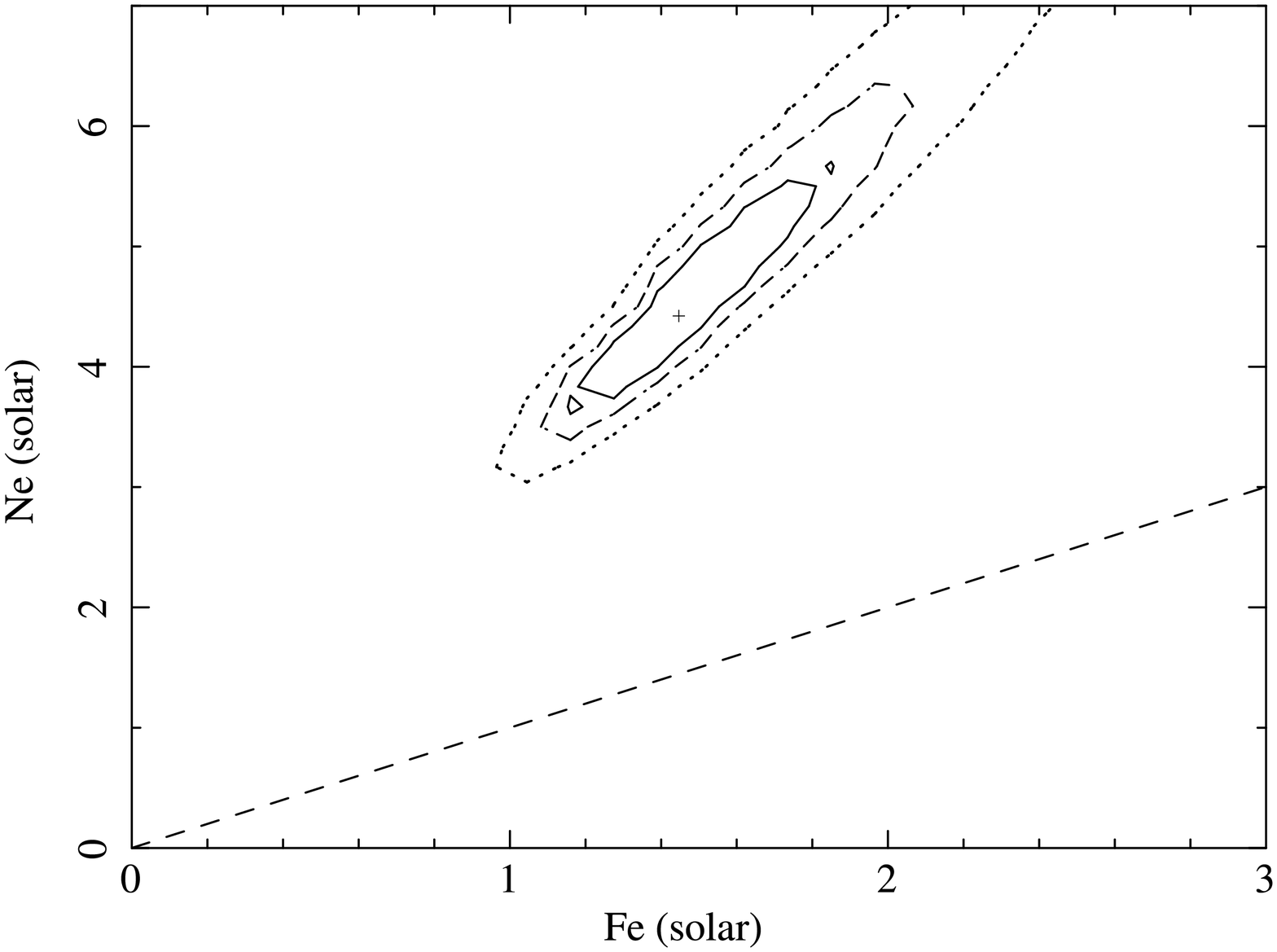}
\plottwo{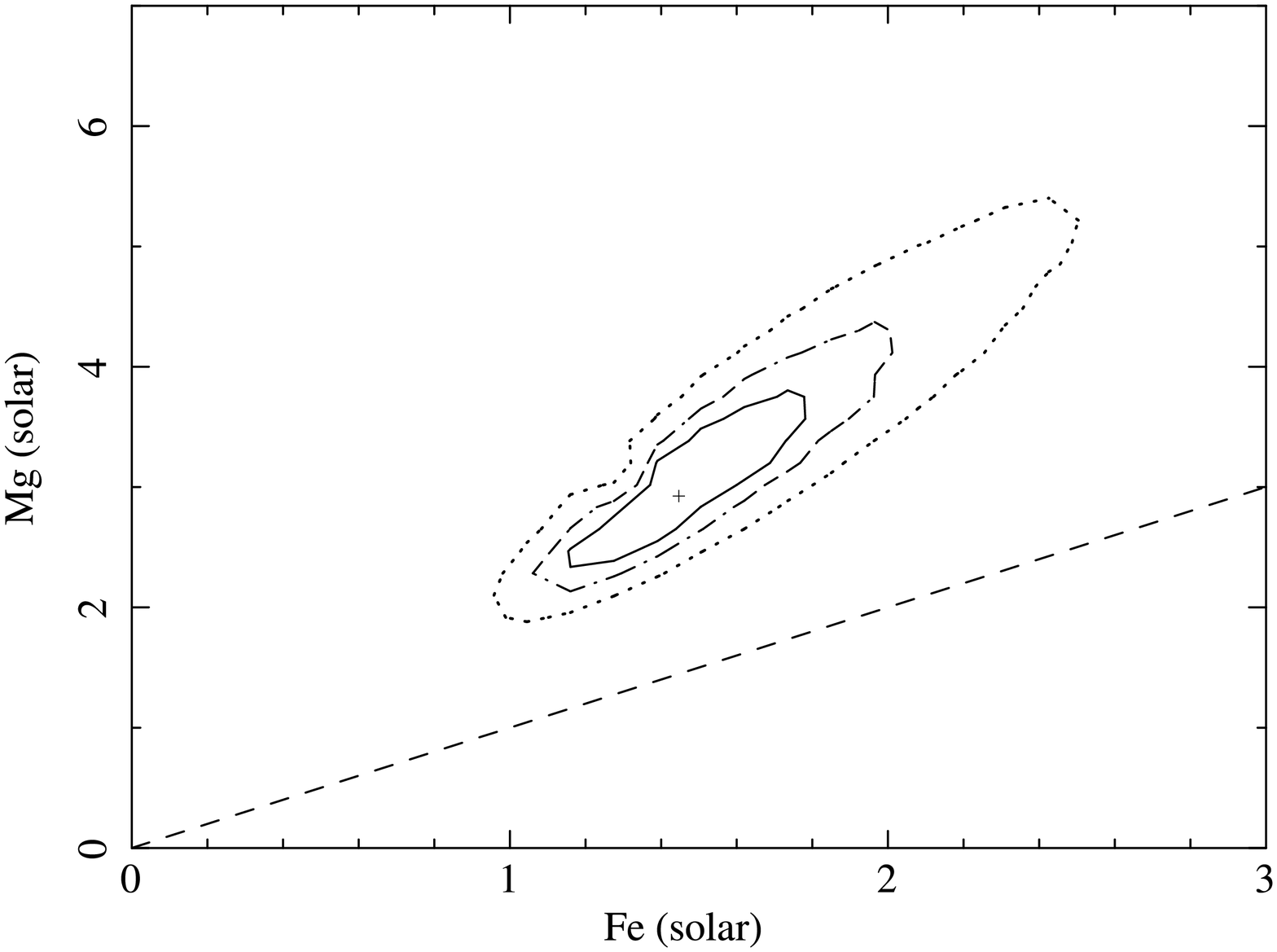}{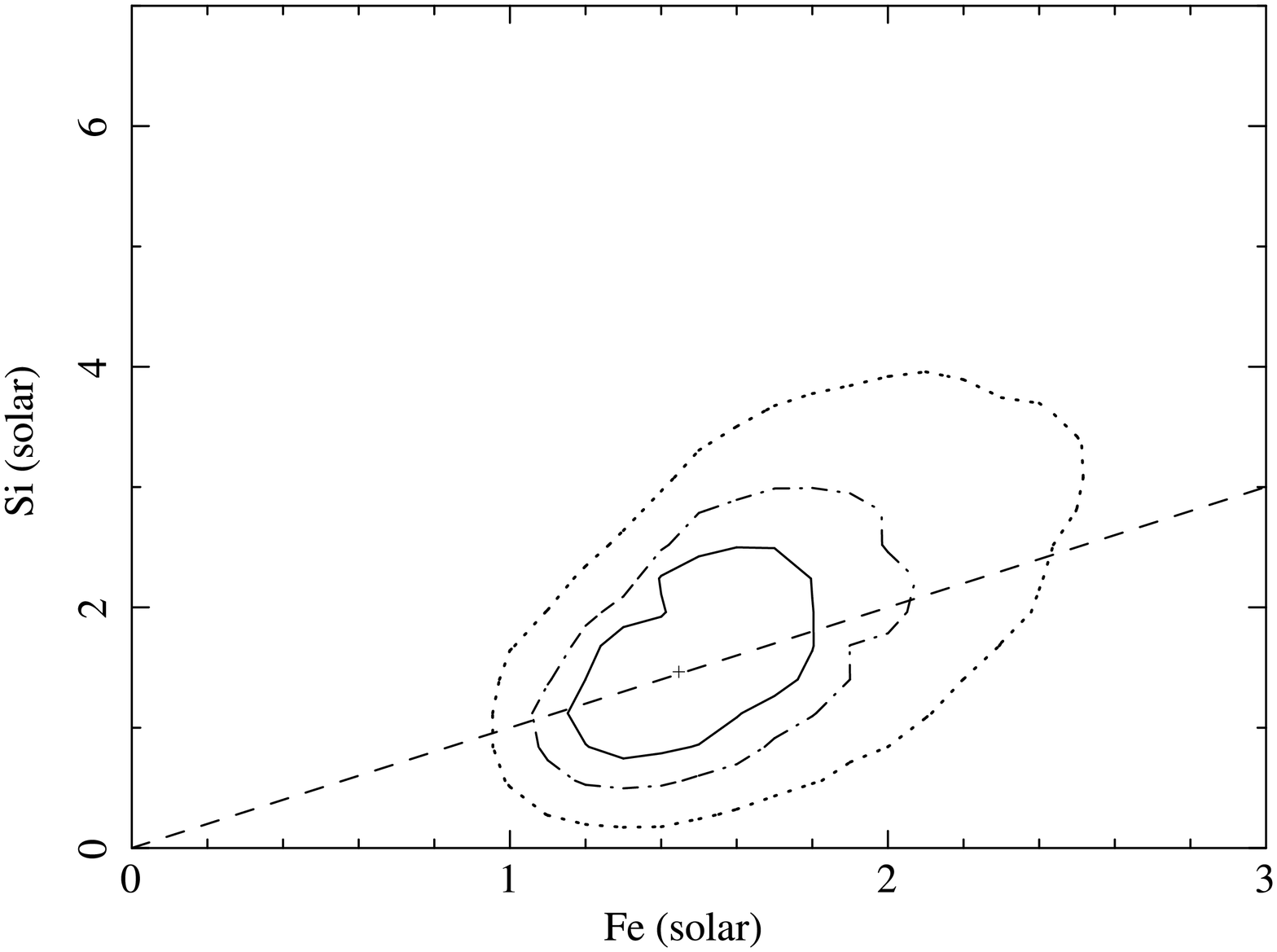}
\caption{Confidence contours for the abundance ratios of O, Ne, Mg, 
  and Si to Fe, determined from the Head spectrum. The solid, dashed, 
  and dotted contours are given for confidence levels of 68\%, 90\%, 
  and 99\%, respectively. The solar abundance ratios are shown by
  the dashed lines. 
  \label{fig:cont_abun}}
\end{figure*}

\subsubsection{Tail, NW, and SE regions}
\label{sssec:other}

We next analyzed the spectra of the Tail region 
using the same model and assumption as for the Head region. 
We found that the best-fit result required a low 
interstellar absorption of $N_{\rm H} < 8.0 \times 10^{19}$~cm$^{-2}$ 
(the best-fit value was pegged at zero). This may be due to 
incomplete calibration information for the contamination layer on 
the OBF at the large off-axis angle region (Koyama et al.\ 2007). 
We thus fixed the column density to the best-fit value of the Head 
region ($N_{\rm H} = 1.3 \times 10^{20}$~cm$^{-2}$) and fitted again. 
We obtained an acceptable result ($\chi ^2$/d.o.f.\ = 208/168)
as given in Table~\ref{tab:fit}. 
We note that all the parameter values are fully consistent with those
derived for the case of $N_{\rm H}$ was freely fitted.
The elemental abundances of the Tail region are found to be significantly 
lower than those of the Head region, except for Si. If we fit the Tail
spectrum fixing the abundances to the best-fit 
values for the Head region given in Table~\ref{tab:fit}, 
the fit was statistically worse ($\chi ^2$/d.o.f.\ = 270/173) based on 
an F-test (probability of $\sim 2 \times 10^{-8}$), and thus was rejected. 
We therefore conclude that the abundances of some elements at the Tail 
region are really depleted relative to those at the Head region. 


The spectra of the NW and SE regions were also fitted with the 
$N_{\rm H}$ fixed.  The best-fit parameters 
(listed in Table~\ref{tab:fit}) were found to be consistent with 
those of the Head region, so that the elemental abundances are
consistent with the values for the Head. Freely fitting the 
$N_{\rm H}$ did not change these results, although the best-fit 
$N_{\rm H}$ values were unreasonably low probably due to the same 
reason seen in the Tail spectrum.

\section{Discussion}
\label{sec:discussion}

\subsection{Chemical Composition}
\label{ssec:chemical}

We have measured enhanced abundances of the heavy elements in 
Vela shrapnel~B, for the first time, strongly supporting that 
shrapnel~B is an ejecta fragment (originating from the explosion 
of the Vela SN), similar to shrapnels~A and D (KT05; KT06). 
We note, however, that it is difficult to determine the absolute 
values of the elemental abundances, because the bremsstrahlung
associated with high-$Z$ ions contributes significantly 
to the continuum emission. Indeed, if we fix the O abundance  
to be 10 times solar and fit the Head spectrum, we
obtain an acceptable $\chi ^2$/d.o.f.\ value of 409/319.

The derived relative abundances among the heavy 
elements are much more reliable, since they are mainly determined 
from the intensity ratios of the emission lines. 
Figure~\ref{fig:cont_abun} shows the confidence contours for the 
abundances of O, Ne, Mg, and Si against that of Fe, derived from 
the spectrum of the Head region. We can see that the abundance ratios 
of Ne/Fe and Mg/Fe are enhanced over the solar values. 
Furthermore, the O/Fe ratio is also higher than unity at $>$ 99\%
confidence.  These results match the nucleosynthesis  
yields of core-collapse SNe (e.g., Thielemann et al.\ 1996) rather than 
thermonuclear Type Ia SNe (e.g., Iwamoto et al.\ 1999). 
The obtained relative abundances (with 90\% error) for the Head region 
are listed in Table~\ref{tab:abundance}, along with those for 
shrapnels~A and D (KT05; KT06), for comparison. 
Shrapnel~A is obviously dominated by Si, while shrapnel~D 
is abundant in O, Ne, and Mg, which are synthesized exterior 
to the Si-rich layer of the progenitor. 
The abundance pattern of shrapnel~B is relatively similar to those 
found in shrapnel~D, except that the over-abundances of the lighter elements 
are less prominent. 
This suggests more significant mixing with the interstellar 
medium (ISM) in shrapnel~B than in shrapnel~D. The same interpretation 
has been advanced argued for some ejecta knots discovered around the northeast 
rim of the Vela SNR (Miceli et al.\ 2008) that have abundance patterns  
very similar to those of shrapnel~B.

We find that the absolute abundances of O, Ne, and Mg in the Tail region 
are significantly lower than those in the Head and the other regions. 
This suggests that the contribution of the ISM is larger at the Tail 
region than elsewhere. 
Note that the FOV of the {\it Suzaku} observation covers a very limited 
portion of the shrapnel (see Figure~\ref{fig:rosat}). In order to 
investigate the more detailed spatial distribution of the elemental 
abundances, we need to observe a larger area of shrapnel~B. 
This will help us to separate the ejecta and ISM components more clearly.

\begin{table}
\begin{center}
\caption{Abundance ratios of Vela shrapnels.\tablenotemark{a}
  \label{tab:abundance}}
\begin{tabular}{lccc}
\tableline\tableline
  ~ & Shrapnel B\tablenotemark{b} & Shrapnel A\tablenotemark{c} & Shrapnel D\tablenotemark{d}\\
\tableline
  O/Fe  &  1.2 (1.1--1.3)  &  0.4  &  4.6  \\
  Ne/Fe &  3.1 (2.8--3.5)  &  1.1  &  9.8  \\
  Mg/Fe &  2.1 (1.7--2.4)  &  1.0  &  10   \\
  Si/Fe &  1.1 (0.4--1.8)  &  3.8  &  0.7  \\
  \tableline
\end{tabular}
\tablenotetext{a}{Abundance ratios are shown in terms of the solar values 
   from Anders \& Grevesse (1989).}
\tablenotetext{b}{The values for the Head region.}
\tablenotetext{c}{Katsuda \& Tsunemi (2006).}
\tablenotetext{d}{Katsuda \& Tsunemi (2005); Note that the Si abundance is assumed to be solar.}
\end{center}
\end{table}

\subsection{Plasma Properties}
\label{ssec:plasma}

The spectra from all the regions are found to be reproduced well by a
NEI plasma model with a single electron temperature of $kT_e \sim 0.6$~keV. 
The obtained $n_et$ value is on the order of $10^{10}$~cm$^{-3}$~s, which 
indicates that the plasma is far from CIE. 
The {\it ROSAT} study showed that the 
electron temperature of shrapnel~B was $kT_e = 0.39 \pm 0.12$~keV 
(Aschenbach et al.\ 1995), which is significantly lower than our result. 
The {\it ASCA} study also derived a lower temperature of 
$kT_e = 0.33 \pm 0.01$~keV (Tsunemi et al.\ 1999b). 
These inconsistencies probably arise because the previous studies applied 
the CIE plasma model. Indeed, we obtain a similarly low temperature from 
the {\it Suzaku} spectrum using CIE models (see \S\ref{sssec:head}). 
The sum of the luminosities of the four regions is 
$L_{\rm X} \sim 3.4 \times 10^{31}$~ergs~s$^{-1}$ in the 0.2--2.0~keV band. 
The {\it ROSAT} PSPC image (Figure~\ref{fig:rosat}) shows that the flux 
in the XIS FOV is $\sim 25\%$ of that of the entire shrapnel so that
the total luminosity of shrapnel~B is estimated to be  
$L_{\rm X} \sim 1.3 \times 10^{32}$~ergs~s$^{-1}$.

Assuming an axially symmetric structure for the shrapnel, the emission 
depth of the plasma is roughly estimated to be $8 \times 10^{18}$~cm 
at a distance of 250~pc. Therefore, using the best-fit elemental 
abundances given in Table~\ref{tab:fit}, the EM of the Head and Tail 
regions correspond to the proton densities 
of 0.06~cm$^{-3}$ and 0.05~cm$^{-3}$, respectively. 
(A homogeneous density distribution is also assumed here.) 
These values are lower than the typical density of shrapnel~D 
($\sim 0.2$~cm$^{-3}$: KT05), which is generally consistent with 
our interpretation that shrapnel~B is more significantly mixed with the 
ISM than is shrapnel~D. For the Head region, the plasma mass is estimated 
to be $1.5\times 10^{-3}~M_{\odot}$, while the mass of the heavy elements 
is $5.2\times 10^{-5}~M_{\odot}$. 
Assuming a volume for the entire shrapnel of $5\times 10^{56}$~cm$^3$ 
and a uniform density of 0.05~cm$^{-3}$, the total mass in 
shrapnel~B is $\sim 0.02~M_{\odot}$. This is comparable to that of 
shrapnel~A (Tsunemi et al.\ 1999a; KT06), although shrapnel~B's
angular size is much larger.  
We note, however, that the plasma mass depends on the absolute 
abundances of heavy elements, which are difficult to determine
the metal-rich plasma such as those in the Head region 
(as mentioned in \S\ref{ssec:chemical}). 
If we assume that the plasma is composed purely of 
metals (without H and He), the mass of the Head region decreases to 
$3.6\times 10^{-4}~M_{\odot}$.

For shrapnels~A and D, the electron temperatures were found to decrease 
gradually toward the trailing regions of the fragments (KT05; 06).  
If the shrapnels are simply breakout features out of the blast wave, 
the temperature observed is expected to increase toward the center of 
the explosion. Therefore, KT06 argued that the higher temperature at 
the head regions supports the ejecta origin of the shrapnels. 
However, we do not find any significant temperature variation between 
the Head and Tail regions of shrapnel~B. 
Follow-up observations of this shrapnel with a deep exposure and a large 
FOV are required to conclusively reveal the $kT_e$ distribution.

\section{Summary}
\label{sec:summary}

We have analyzed {\it Suzaku} data of Vela shrapnel~B. 
Utilizing the good sensitivity and spectroscopic performance of the XIS, 
we clearly detect several emission line blends from highly ionized
heavy elements.  The spectrum of the Head region is well fitted with
an NEI plasma with an electron temperature of $kT_e \sim 0.6$~keV. The
abundances of some heavy elements are found to be significantly
enhanced above the solar values. This result strongly suggests that
shrapnel~B originates from the ejecta of the Vela SN.  The abundances of 
O, Ne, and Mg relative to Fe are higher than 
the solar ratios. Shrapnel~B can therefore be identified with the
outer layer of the progenitor, similar to shrapnel~D.  
However, the over-abundances of the lighter elements are less prominent 
than for shrapnel~D, indicating that shrapnel~B has been 
mixed with a larger mass of ISM than shrapnel~D. 
The spectrum from the Tail region is reproduced by 
a similar model as the Head region, but with significantly lower 
absolute abundances of O, Ne, and Mg. 
This indicates a relatively large contribution from swept-up ISM for
the Tail region. No significant temperature variation 
between the Head and the Tail regions is found. 
The total luminosity (in 0.2--2.0~keV band) and mass of the entire 
shrapnel are estimated to be 
$L_{\rm X} \sim 1.3 \times 10^{32}$~ergs~s$^{-1}$, and 
$\sim 0.02~M_{\odot}$, respectively. 

\acknowledgments

The authors deeply appreciate helpful comments on revising the 
manuscript from Dr.\ Una Hwang. A number of constructive suggestions 
from the referee greatly helped us to improve the quality of the paper. 
We also thank Dr.\ Junko S.\ Hiraga for useful discussions. H.Y.\ is 
supported by the Special Postdoctoral Researchers Program in RIKEN. S.K.\ 
is a Research Fellow of Japan Society for Promotion of Science (JSPS). 






\begin{thebibliography}{}

\bibitem[Anders \& Grevesse(1989)]{1989GeCoA..53..197A} Anders, E., \& 
Grevesse, N.\ 1989, \gca, 53, 197 

\bibitem[Aschenbach et al.(1995)]{1995Natur.373..587A} Aschenbach, B., 
Egger, R., \& Trumper, J.\ 1995, \nat, 373, 587 


\bibitem[Borkowski et al.(2001)]{2001ApJ...548..820B} Borkowski, K.~J., 
Lyerly, W.~J., \& Reynolds, S.~P.\ 2001, \apj, 548, 820 

\bibitem[Brickhouse et al.(2000)]{2000ApJ...530..387B} Brickhouse, N.~S., 
Dupree, A.~K., Edgar, R.~J., Liedahl, D.~A., Drake, S.~A., White, N.~E., 
\& Singh, K.~P.\ 2000, \apj, 530, 387 

\bibitem[Cha et al.(1999)]{1999ApJ...515L..25C} Cha, A.~N., Sembach, K.~R., 
\& Danks, A.~C.\ 1999, \apjl, 515, L25 

\bibitem[Decourchelle et al.(2001)]{2001A&A...365L.218D} Decourchelle, A., 
et al.\ 2001, \aap, 365, L218 

\bibitem[Fesen et al.(2006)]{2006ApJ...636..859F} Fesen, R.~A., et al.\ 
2006, \apj, 636, 859 

\bibitem[Gvaramadze(1999)]{1999A&A...352..712G} Gvaramadze, V.\ 1999, 
\aap, 352, 712 

\bibitem[Iwamoto et al.(1999)]{1999ApJS..125..439I} Iwamoto, K., Brachwitz, 
F., Nomoto, K., Kishimoto, N., Umeda, H., Hix, W.~R., 
\& Thielemann, F.-K.\ 1999, \apjs, 125, 439 

\bibitem[Katsuda et al.(2008)]{2008ApJ...678..297K} Katsuda, S., Mori, K., 
Tsunemi, H., Park, S., Hwang, U., Burrows, D.~N., Hughes, J.~P., 
\& Slane, P.~O.\ 2008, \apj, 678, 297 

\bibitem[Katsuda \& Tsunemi(2006)]{2006ApJ...642..917K} Katsuda, S., \& 
Tsunemi, H.\ 2006, \apj, 642, 917 

\bibitem[Katsuda \& Tsunemi(2005)]{2005PASJ...57..621K} Katsuda, S., \& 
Tsunemi, H.\ 2005, \pasj, 57, 621 

\bibitem[Koyama et al.(2007)]{2007PASJ...59S..23K} Koyama, K., et al.\ 
2007, \pasj, 59, 23 


\bibitem[Laming \& Hwang(2003)]{2003ApJ...597..347L} Laming, J.~M., \& 
Hwang, U.\ 2003, \apj, 597, 347 

\bibitem[Lu \& Aschenbach(2000)]{2000A&A...362.1083L} Lu, F.~J., \& 
Aschenbach, B.\ 2000, \aap, 362, 1083 

\bibitem[Miceli et al.(2008)]{2008ApJ...676.1064M} Miceli, M., Bocchino, 
F., \& Reale, F.\ 2008, \apj, 676, 1064 


\bibitem[Miyata et al.(2001)]{2001ApJ...559L..45M} Miyata, E., Tsunemi, H., 
Aschenbach, B., \& Mori, K.\ 2001, \apjl, 559, L45 

\bibitem[Park et al.(2007)]{2007ApJ...670L.121P} Park, S., Hughes, J.~P., 
Slane, P.~O., Burrows, D.~N., Gaensler, B.~M., 
\& Ghavamian, P.\ 2007, \apjl, 670, L121 

\bibitem[Serlemitsos et al.(2007)]{2007PASJ...59S...9S} Serlemitsos, P.~J., 
et al.\ 2007, \pasj, 59, 9 

\bibitem[Smith et al.(2001)]{2001ApJ...556L..91S} Smith, R.~K., Brickhouse, 
N.~S., Liedahl, D.~A., \& Raymond, J.~C.\ 2001, \apj, 556, L91 

\bibitem[Taylor et al.(1993)]{1993ApJS...88..529T} Taylor, J.~H., 
Manchester, R.~N., \& Lyne, A.~G.\ 1993, \apjs, 88, 529 

\bibitem[Thielemann et al.(1996)]{1996ApJ...460..408T} Thielemann, F.-K., 
Nomoto, K., \& Hashimoto, M.-A.\ 1996, \apj, 460, 408 

\bibitem[Tsunemi et al.(1999a)]{1999PASJ...51..711T} Tsunemi, H., Miyata, 
E., \& Aschenbach, B.\ 1999a, \pasj, 51, 711 

\bibitem[Tsunemi et al.(1999b)]{1999hxra.conf..310T} Tsunemi, H., Miyata, 
E., Hiraga, J., \& Aschenbach, B.\ 1999b, Highlights in X-ray Astronomy, 310 

\bibitem[Uchiyama et al.(2008)]{2008arXiv0810.0873U} Uchiyama, H., et al.\ 
2008, arXiv:0810.0873 (accepted by \pasj)

\bibitem[Weiler \& Sramek(1988)]{1988ARA&A..26..295W} Weiler, K.~W., \& 
Sramek, R.~A.\ 1988, \araa, 26, 295 

\bibitem[Winkler \& Kirshner(1985)]{1985ApJ...299..981W} Winkler, P.~F., \& 
Kirshner, R.~P.\ 1985, \apj, 299, 981 




\end{thebibliography}
\end{document}